\documentclass[aps,twocolumn]{revtex4-1}
\usepackage{amsmath,amssymb}
\usepackage{graphicx}
\usepackage{tikz}
\usepackage{bm}
\usepackage[T1]{fontenc}
\usepackage[utf8]{inputenc}
\usepackage{geometry}
\usepackage[colorlinks,linkcolor=blue,citecolor=blue]{hyperref}
\usepackage{subfig}

\newcommand{\be}{\begin{equation}}
\newcommand{\ee}{\end{equation}}
\newcommand{\bea}{\begin{eqnarray}}
\newcommand{\eea}{\end{eqnarray}}

\begin{document}



\title{String-based parametrization of nucleon GPDs at any skewness: \\a comparison to lattice QCD}

\author{Kiminad A. Mamo}
\email{kamamo@wm.edu}
\affiliation{
Physics Department, William \& Mary, Williamsburg, VA 23187, USA}

\author{Ismail Zahed}
\email{ismail.zahed@stonybrook.edu}
\affiliation{Center for Nuclear Theory, Department of Physics and Astronomy, Stony Brook University, Stony Brook, New York 11794-3800, USA}

\date{\today}

\begin{abstract}


We introduce a string-based parametrization for nucleon quark and gluon generalized parton distributions (GPDs) valid at all skewness values. The conformal moments of the GPDs are expressed as sums of the spin-j nucleon A-form factor, and the skewness-dependent spin-j nucleon D-form factor. This representation, which fulfills the polynomiality condition (due to Lorentz invariance) and does not rely on model-specific assumptions, is derived from t-channel string exchanges in AdS spaces. The spin-j nucleon D-form factor is closely related to the the spin-j nucleon A-form factor. We use the Mellin moments from empirical parton distributions to model the spin-j nucleon A-form factors. Using only five Regge slope parameters, fixed from the electromagnetic and gravitational form factors, our string-based parametrization  generates accurate singlet, non-singlet, isovector, and flavor-separated nucleon quark GPDs, along with symmetric nucleon gluon GPDs from their Mellin-Barnes integral representations. Our isovector nucleon quark GPD is in agreement  with existing lattice data. Our string-based parametrization should facilitate the empirical extraction and global analysis of nucleon GPDs in exclusive processes, bypassing the deconvolution challenge.

\end{abstract}

\maketitle

\section{Introduction}
\label{Introduction}
On the light front  hadrons are assemblies of valence and sea partons. These constituents, directly probed through hard inclusive and semi-inclusive processes, owe their "visibility" to relativistic time dilation and  asymptotic freedom~\cite{Callan:1969uq,Gross:1973id,Politzer:1973fx}. This allows for the separation of hard processes into calculable perturbative factors, and non-perturbative matrix elements like parton distribution functions (PDFs) and fragmentation functions (FFs), the central tenets in modern high-energy scattering data~\cite{Collins:1989gx,Collins:2011zzd}.

At the center of our investigation are the Generalized Parton Distributions (GPDs), extended PDFs as off-forward matrix elements of leading twist QCD operators in highly boosted hadron states~\cite{Muller:1994ses,Radyushkin:1996nd,Radyushkin:1996ru,Ji:1996nm,Radyushkin:1997ki,Ji:1998pc,Goeke:2001tz}, see~\cite{Diehl:2003ny,Belitsky:2005qn,Mezrag:2022pqk} for review. These distributions are inherently non-perturbative. They offer a multi-dimensional view of nucleons, describing  their spin, mass and mechanical structure via the variables $x$, $\eta$, and $t$—respectively representing the parton momentum fraction, skewness, and total momentum transfer~\cite{Diehl:2003ny,Belitsky:2005qn,Mezrag:2022pqk}. 

Despite their richness, GPDs elude easy probing through first principle QCD lattice simulations, owing to their inherent light cone structure. Yet, progress has been recently made  by recent lattice QCD collaborations~\cite{Chen:2019lcm,Lin:2023gxz,Lin:2023kxn,Lin:2020rxa,Lin:2021brq,Alexandrou:2020zbe,Alexandrou:2021bbo,Bhattacharya:2022aob}, using the theoretical considerations proposed initially in~\cite{Ji:2013dva,Ji:2014gla,Ji:2020ect}, and since extended by others~\cite{Radyushkin:2017cyf,Orginos:2017kos}. A number of experiments at various electron facilities~\cite{dHose:2016mda} are currently under way to shed more light on the phenomenology of GPDs~\cite{Guidal:2013rya,Kumericki:2016ehc}, with more systematic exploration anticipated at the forthcoming EIC (electron-ion collider) facility~\cite{Accardi:2012qut,AbdulKhalek:2021gbh,Anderle:2021wcy}.


GPDs at finite skewness, have been modeled in various theoretical frameworks, owing to their central  role in describing the internal structure of hadrons. These include standard quark models~\cite{Ji:1997gm,Anikin:2001zv,Scopetta:2002xq,Boffi:2002yy,Scopetta:2003et}, chiral soliton models~\cite{Petrov:1998kf,Penttinen:1999th,Goeke:2001tz}, AdS/QCD models~\cite{Vega:2010ns,Rinaldi:2017roc,Traini:2016jko}, light front and holography hybrid models~\cite{deTeramond:2018ecg,Mondal:2015uha,Chakrabarti:2015ama,Mondal:2017wbf,Chakrabarti:2013gra,Gurjar:2022jkx}, light front models~\cite{Shuryak:2023siq,Liu:2024umn}, double distribution models~\cite{Radyushkin:1998es,Radyushkin:1998bz,Polyakov:1999gs,Musatov:1999xp,Goloskokov:2009ia,Goloskokov:2005sd,Goloskokov:2007nt,Goeke:2001tz,Vanderhaeghen:1998uc,Vanderhaeghen:1999xj,Guidal:2004nd}, conformal moment based models at small skewness~\cite{Polyakov:1998ze,Polyakov:2002wz,Mueller:2005ed,Kumericki:2007sa,Kumericki:2009uq,Polyakov:2009xir,Muller:2014wxa,Cuic:2024fum,Guo:2022upw,Guo:2023ahv,Cuic:2023mki}, and recently using lattice data for Compton form factors~\cite{Hannaford-Gunn:2024aix}. Specifically, \cite{Vega:2010ns} and \cite{deTeramond:2018ecg} use an integral representation of only the \( j=1 \) quark conformal moment, which corresponds to the electromagnetic form factors, hence they are restricted to zero skewness. \cite{Mondal:2015uha} addresses non-zero skewness but is limited to the DGLAP region (specifically as \( x \rightarrow 1 \)), and therefore does not satisfy the polynomiality condition because it does not cover the full integral range of \( x \).

The unpolarized GPD $H(x,\eta,t)$  and polarized GPD $\tilde{H}(x,\eta,t)$, are generalized form factors.
These distributions reduce to ordinary PDFs in the forward limit, and turn into pertinent Compton form factors when summed over $x$. The inherent complexity and non-uniqueness in the deconvolution of GPDs from the Compton form factors of deep virtual Compton scattering (DVCS) and deep virtual meson production (DVMP), presents a significant theoretical challenge for their model independent extraction from data~\cite{Bertone:2021yyz}.
One approach to bypass the deconvolution problem, is to parametrize the conformal moments of GPDs \cite{Guo:2022upw,Guo:2023ahv}, which are highly constrained by Lorentz invariance (polynomiality condition), and can be interpreted in terms of t-channel spin-$j$ resonances \cite{Polyakov:1998ze,Polyakov:2002wz}.

For a fixed Mandelstam $t$ and positive skewness $\eta$, GPDs exhibit distinct behaviors in two kinematic regimes: the DGLAP regime for $|x| > \eta$ and the ERBL regime for $|x| < \eta$. In the DGLAP regime, the dynamics correspond to either quark or antiquark momentum redistribution, while in the ERBL regime, it is akin to the emission of meson-like quark-antiquark pairs~\cite{Diehl:2003ny,Belitsky:2005qn}.

We propose to use  the holographic string-based approach (see the review~\cite{Nastase:2015wjb}, and references therein), to leverage the holographic principle to bypass the de-convolution issue entirely. By employing Gegenbauer moments with fixed conformal spin-j, each moment is directly mapped to its gravity dual, allowing a detailed reconstruction of the GPDs in the ERBL regime, as we noted recently~\cite{Mamo:2022jhp}. This method not only simplifies the encoding of stringy properties of QCD at low resolution, but also aligns with empirical Regge phenomenology and Veneziano amplitudes, capturing critical aspects of QCD dispersive analysis such as crossing symmetry, unitarity, and spectral densities in a framework with remarkably few parameters, e.g., open and closed string intercepts and slopes. The GPDs are then expressed through an expansion in integer-valued spin-n conformal moments of quark and gluon PDFs, structured using an orthogonal basis of conformal partial waves (PWs). By employing the Sommerfeld-Watson transform, this series is transformed into an integral over analytically continued complex spin-j conformal moments, which preserves analyticity across all regions of skewness~\cite{Mueller:2005ed}.

The outline of the paper is as follows: Section~\ref{sec_2} elaborates on the expansion of quark and gluon GPDs in conformal PWs, particularly in the ERBL region, and discusses their analytical continuation to DGLAP region. Section~\ref{sec_3} details the holographic string parametrization of the conformal moments used for reconstructing the quark and gluon GPDs across all values of parton-x. Section~\ref{sec_4} compares the conformal moments and the reconstructed GPDs to those derived from Euclidean lattice simulations within the large momentum effective theory (LaMET). Finally, Section~\ref{sec_5} summarizes our findings and conclusions. The appendix provides an overview of the empirical nucleon MSTW PDFs used in our analysis.


\section{Quark and gluon GPDs}
\label{sec_2}
A natural organization of  the quark and gluon 
GPDs in the ERBL regime, is a series expansion in terms of the conformal
partial waves (PWs), with the conformal moments as coefficients. The conformal PWs are  Gegenbauer polynomials in  parton-x of the struck quark in the GPD, that diagonalize the pertinent GPD evolution operator. 
The conformal moments carry the essential dynamics of the quarks and gluons in hadrons.

The extension to the DGLAP  regime requires analytical continuation of the
series expansion, using a Sommerfeld-Watson transform. The result is a GPD given as an integral transform of spin-j conformal moments, following a pertinent analytical continuation of the PWs. In this section we recall the essential relations needed in this continuation,  for both the quarks and gluons.


\subsection{Quark GPDs}
The quark GPDs, defined as $H_q(x, \eta, t; \mu)$, can be naturally expanded in terms of Gegenbauer polynomials in parton-x. These polynomials are best structured in terms of PWs which are eigenstates of the pertinent QCD evolution operator for a given GPD.

\subsubsection{Series expansion}
More specifically, a generic quark GPD admits a series expansion for $|x|<\eta$
\begin{eqnarray}
\label{ExpansionGPDeigenfunctionsH222VectorQuark}
H_q(x, \eta, t; \mu) = \sum_{n=1}^{\infty} (-1)^{n-1} p_n(x,\eta) \mathbb{F}_q(n, \eta, t; \mu),
\end{eqnarray}
with the conformal moments $\mathbb{F}_q(n, \eta, t; \mu)$ as coefficient
functions. The conformal partial waves (PWs), $p_n(x, \eta)$, form an orthonormal basis set defined as
 \begin{widetext}
\begin{eqnarray}
\label{Def-pn}
p_n(x, \eta) = \frac{1}{\eta^n} p_n\left(\frac{x}{\eta}\right) \quad \mbox{with} \quad p_n(x) = \theta(1-|x|) \frac{2^{n-1} \Gamma(3/2 + n)}{\Gamma(3/2)\Gamma(2 + n)} (1 - x^2) C_{n-1}^{3/2}(-x),
\end{eqnarray}
\end{widetext}
where $C_n^{\nu}(x)$ are Gegenbauer polynomials.
The PWs encode reflection symmetry $p_n(-x,\eta)=(-1)^np_n(x, \eta)$,
and turn to distributions at zero skewness
\begin{equation}
p_n(x, \eta = 0) = \frac{1}{n!}\delta^{(n)}(x).
\end{equation}
The orthonormality relation of Gegenbauer polynomials, key
to the inversion of the GPDs, is captured by 
\begin{eqnarray}
\label{OrtRel} \int_{-1}^1 dx\, c_{n'}(x, \eta) p_n(x, \eta) = (-1)^{n-1} \delta_{n'n}\,,
\end{eqnarray}
where
\bea
\label{Def-c}
 c_n(x,\eta) &=&  \eta^{n-1}\times c_j\!\left(\!\frac{x}{\eta}\!\right)
\nonumber\\
c_n(x) &=&
\frac{\Gamma(3/2)\Gamma(n)}{2^{n-1} \Gamma(1/2+n)}
\times C_{n-1}^{3/2}\left(x\right)\,,
\eea
Note that the conformal moments are Gegenbauer moments of the GPD
\bea
\label{QuarkConfMomentsALL}
\mathbb{F}_q(n,\eta, t  ; \mu^2)
=\int_{-1}^1 dx \  c_{n}\left(x,\eta\right)
H_q (x, \eta, t  ; \mu)\, ,
\eea
for $n=1,2,...$. Also note that in the forward limit, the spin-n conformal moments of the quark GPDs reduce to the spin-n Mellin moments of the quark GPDs, since
\begin{eqnarray}
\label{Def-csca} \lim_{\eta\to 0} c_n(x,\eta) = x^{n-1}\,.
\end{eqnarray}

\subsubsection{Non-singlet (Valence) Quark GPDs}
The non-singlet or valence quark GPDs, denoted as $H^{(-)}_q(x, \eta, t; \mu)$, represent the difference between quark distributions and their antiparticle counterparts within the hadron. This difference is key for understanding the valence quark structure. They are defined as
\begin{widetext}
\begin{eqnarray}
\label{ExpansionGPDValenceVector}
H^{(-)}_q(x, \eta, t; \mu) = H_q(x, \eta, t; \mu) + H_q(-x, \eta, t; \mu) = \sum_{n=1}^{\infty} (-1)^{n-1} \left(p_n(x,\eta) + p_n(-x,\eta)\right) \mathbb{F}_q^{(-)}(n, \eta, t; \mu),
\end{eqnarray}
\end{widetext}
for odd $n=1,3,\ldots$, with support $0\leq x\leq\eta$. (\ref{ExpansionGPDValenceVector}) can be extended 
to the complex j-plane with the help of the  Sommerfeld-Watson integral transform, with support $0\leq x\leq1$, as
\bea
\label{HvalenceFinal}
&&H^{(-)}_q(x, \eta, t; \mu) = \nonumber\\
&&\frac{1}{2i}\int_{\mathbb{C}} dj\, \frac{1}{\sin(\pi j)} (p_j(x,\eta) + p_j(-x,\eta)) \mathbb{F}_q^{(-)}(j, \eta, t; \mu),\nonumber\\
\eea
by extending spin-n to complex spin-j, as suggested in~\cite{Mueller:2005ed} (for recent discussions see~\cite{,Guo:2022upw} and references therein). Note that the factor $\frac{1}{\sin(\pi j)}$, with residues $\mbox{Res}_{j=n}\, 1/\sin(\pi  j) = (-1)^{n-1}/\pi$ for $n={1,3,\cdots}$, yields the original series,
barring additional singularities within the integration contour $\mathbb{C}$.

The analytic continuation of the PWs $p_n(x, \eta)$ to the complex $j$-plane, can be done using their Schl{\"a}fli integral representation~\cite{Mueller:2005ed}
\begin{eqnarray}
\label{Def-p-allhere}
&&p_j(x,\eta) = \nonumber\\
&&\theta(\eta - |x|) \frac{1}{\eta^j} \mathcal{P}_j\left(\frac{x}{\eta}\right) + \theta(x - \eta) \frac{1}{x^j} \mathcal{Q}_j\left(\frac{x}{\eta}\right),
\end{eqnarray}
The functions \(\mathcal{P}_j\) and \(\mathcal{Q}_j\) are hypergeometric functions
\begin{widetext}
\bea
\label{Def-p-P}
\mathcal{P}_j\left(\frac{x}{\eta}\right) &=&
\left(1+\frac{x}{\eta}\right) {_2F_1}\left(-j, j+1,2\bigg|\frac{1}{2}\left(1+\frac{x}{\eta}\right)\right) \times \frac{2^{j}\Gamma(3/2+j)}{\Gamma(1/2)\Gamma(j)},\nonumber\\
\label{Def-p-Q}
\mathcal{Q}_j\left(\frac{x}{\eta}\right) &=& {_2F_1}\left(\frac{j}{2},\frac{j+1}{2}; \frac{3}{2} + j\bigg|\frac{\eta^2}{x^2}\right) \times \frac{\sin(\pi j)}{\pi},
\eea
\end{widetext}
for all values of parton-x. The complex-valued functions \(\mathcal{P}_j\) cover the ERBL regime, and \(\mathcal{Q}_j\) cover the DGLAP regime. 
This illustrates how the analytic continuation process for the PWs is
captured in the complex spin-j plane.
Note that the functional dependence of the conformal moment $\mathbb{F}_q(j,\eta,t;\mu)$ is common to both regimes, which
makes it accessible from the ERBL regime using dual gravity. This observation is central to our analysis. 
The conformal moments carry the essential hadron dynamics, that interpolate between the 
distribution amplitudes (ERBL regime) and the parton distribution functions (DGLAP regime). 

The quark PWs are continuous across $x=\eta$ since $\mathcal{P}_j\left(1\right)=\mathcal{Q}_j\left(1\right)$, but their slopes (derivatives) are discontinuous.
They reflect on the different physics at work in the two regions. In the ERBL region, 
'wee partons' aggregate into Reggeized ladders (dual to string world sheets), while in the DGLAP region 
at sufficiently large-x, the 'partons' are described by few Fock components. Both descriptions are likely captured by distinct  asymptotic series, leading to (\ref{HvalenceFinal}) by the general structures of analyticity and duality.

To segregate between the contributions stemming from valence and sea quarks partonic distributions in hadrons, we now discuss the singlet and non-singlet  quark GPDs.

\subsubsection{Non-singlet (Isovector) Quark GPDs}
The non-singlet isovector quark GPDs, denoted as $H^{(-)}_{u-d}(x, \eta, t; \mu)$, represents the difference between the up and down quark distributions. They are defined as
\begin{eqnarray}
\label{umdExpansionGPDValenceVector}
H^{(-)}_{u-d}(x, \eta, t; \mu) =\sum_{n=1}^{\infty} (-1)^{n-1} \,p_n(x,\eta)\,\mathbb{F}_{u-d}^{(-)}(n, \eta, t; \mu),\nonumber\\
\end{eqnarray}
for all $n=1,2,3,4,\ldots$. To account for non-integer conformal spins, (\ref{umdExpansionGPDValenceVector}) can be extended 
to the complex j-plane with the help of the  Sommerfeld-Watson integral transform with support $-\eta\leq x\leq1$
\bea
\label{umdHvalenceFinal}
&&H^{(-)}_{u-d}(x, \eta, t; \mu) = \nonumber\\
&&\frac{1}{2i}\int_{\mathbb{C}} dj\, \frac{1}{\sin(\pi j)} \,p_j(x,\eta)\,\mathbb{F}_{u-d}^{(-)}(j, \eta, t; \mu),\nonumber\\
\eea
with the analytically continued conformal PWs $p_j(x,\eta)$ given in (\ref{Def-p-allhere}).

\subsubsection{Singlet (Sea)  Quark GPDs}
The singlet or sea quark GPDs, $H^{(+)}(x, \eta, t; \mu)$, sum  the contributions from all quark flavors,  encompassing both quarks and antiquarks. They reflect on  the sea quark distribution. The singlet combination is defined as
\begin{widetext}
\begin{eqnarray}
\label{ExpansionGPDsinglet222}
H^{(+)}(x, \eta, t; \mu) = \sum_{q=1}^{N_f} H_q(x, \eta, t; \mu) - H_q(-x, \eta, t; \mu) = \sum_{n=2}^{\infty} (-1)^{n-1} (p_n(x,\eta) - p_n(-x,\eta)) \sum_{q=1}^{N_f} \mathbb{F}^{+}_q(j, \eta, t; \mu),\nonumber\\
\end{eqnarray}
for even $n=2,4,\ldots$. Similarly to the non-singlet case, we use the the Sommerfeld-Watson transform 
to extend (\ref{ExpansionGPDsinglet222}) to complex conform spin-j with support $0\leq x\leq1$

\begin{eqnarray}
\label{HseaFinal}
H^{(+)}(x, \eta, t; \mu) = \frac{1}{2i}\int_{\mathbb{C}} dj\, \frac{1}{\sin(\pi j)} (p_j(x,\eta) - p_j(-x,\eta)) \sum_{q=1}^{N_f} \mathbb{F}^{+}_q(j, \eta, t; \mu),
\end{eqnarray}
\end{widetext}

\subsection{Gluon GPDs}
Following our brief discussion of the quark GPDs, we now present the gluon GPDs. Generically, the gluons whether in perturbative or non-perturbative form, are blind to the electric and weak probes, and hence more subtle to quantify. Yet, they  are essential ingredients in the composition of the  mass and  spin of hadrons. 

\subsubsection{Series expansion}
The formulation of the gluon GPDs, $H_g(x, \eta, t; \mu)$, mirrors that  for the quarks, albeit with adaptations to account for the unique characteristics of gluons. The series expansion of the gluon GPDs, in terms of their spin-n conformal moments $\mathbb{F}_g(n, \eta, t; \mu)$, is 
\begin{eqnarray}
\label{ExpansionGPDgluon22Unpol}
H_g(x, \eta, t; \mu) =\sum_{n=2}^{\infty} (-1)^{n+1} {^g\!p}_n(x,\eta) \mathbb{F}_g(n, \eta, t; \mu),\nonumber\\
\end{eqnarray}
with the sum running now unrestricted. 
The gluon PWs ${^g\!p}_n(x, \eta)$, form the basis
functions for the gluon partonic distributions in hadrons
\begin{widetext}
\bea
\label{Def-pnG}
{^g\!p}_n(x, \eta) &=& \frac{1}{\eta^{n-1}} {^g\!p}_n\left(\frac{x}{\eta}\right) \nonumber\\
{^g\!p}_n(x) &=& (-1) \theta(1-|x|) \frac{2^{n-1}3^2\Gamma(3/2+n)}{\Gamma(5/2)\Gamma(3+n)} (1 - x^2)^2 C_{n-2}^{5/2}(-x),
\eea
\end{widetext}
with the reflection symmetry
$${^g\!p}_n(-x, \eta) = (-1)^{n} {^g\!p}_n(x, \eta)\,.$$
The orthogonality relation for the Gegenbauer polynomials, key for the inversion process, is 
\begin{eqnarray}
\label{OrtRelGluon}
\int_{-1}^1 dx\, {^g\!c}_{n'}(x, \eta) {^g\!p}_n(x, \eta) = (-1)^{1+n} \delta_{n'n}\,,
\end{eqnarray}
which ensures that
\bea
\label{GluonConfMomentsALL}
\mathbb{F}^g (n,\eta, t  ; \mu)
=\int_{-1}^1 dx \  {^g\! c}_n(x,\eta)
H^g (x, \eta, t  ; \mu)\, ,
\eea
for $n=2,3,...$, with
\bea
\label{Def-cG} 
{^g\! c}_n(x,\eta) &=&  
\eta^{n-2}\times\, {^g\!c}_j\!\left(\!\frac{x}{\eta}\!\right) \nonumber\\
{^g\!c}_n(x) &=& \frac{\Gamma(5/2)\Gamma(n-1)}{2^{n-2} \Gamma(1/2+n)}\times C_{n-2}^{5/2}(x)\,.
\eea
Note that in the forward limit, the spin-n conformal moments of the gluon GPDs reduce to the spin-n Mellin moments of gluon GPDs since
\begin{eqnarray}
\label{Def-cscaG} \lim_{\eta\to 0} {^g\! c}_n(x,\eta) = x^{n-2}\,.
\end{eqnarray}

\subsubsection{Integral transform}
Similarly to the quark GPDs, the series expansion for the gluon GPD can be analytically continued from real spin-n to complex spin-j using the Sommerfeld-Watson transform
as originally suggested in~\cite{Mueller:2005ed}. The resulting Mellin-Barnes integral representation of the gluon GPD is
\bea
\label{F(x,eta,t)-MBUnpolGluon}
&&H_g(x, \eta, t; \mu) = \nonumber\\
&&\frac{1}{2i} \int_{\mathbb C} dj\, \frac{(-1)}{\sin(\pi j)} {^g\!p}_j(x, \eta) \mathbb{F}_g(j, \eta, t; \mu^2),\nonumber\\
\eea
where the factor $\frac{1}{\sin(\pi j)}$ is included to ensure proper normalization, and the  recovery of the PWs expansion in the absence of other singularities within the contour $\mathbb{C}$. The analytically continued gluon PWs are
\bea
\label{Def-p-allGluonhere}
&&{^g\!p}_j(x, \eta) = \nonumber\\
&&\theta(\eta - |x|) \frac{1}{\eta^{j-1}} {^g\!\cal P}_j\left(\frac{x}{\eta}\right) + \theta(x - \eta) \frac{1}{x^{j-1}} {^g\!\cal Q}_j\left(\frac{x}{\eta}\right),\nonumber\\
\eea
where the functions ${^g\!\cal P}_j$ and ${^g\!\cal Q}_j$ are given by
\begin{widetext}
\bea
\label{Def-p-PGluon}
{^g\!\cal P}_j\left(\frac{x}{\eta}\right) &=& \left(1+\frac{x}{\eta}\right)^2 {_2F_1}\left(-j, j+1, 3 \bigg|\frac{1}{2}\left(1+\frac{x}{\eta}\right)\right) \frac{2^{j-1}\Gamma(3/2+j)}{\Gamma(1/2)\Gamma(j-1)},
\nonumber\\
\label{Def-p-QGluon}
{^g\!\cal Q}_j\left(\frac{x}{\eta}\right) &=& {_2F_1}\left(\frac{j-1}{2}, \frac{j}{2}; \frac{3}{2} + j; \frac{\eta^2}{x^2}\right) \frac{\sin(\pi[j+1])}{\pi}.
\eea\end{widetext}
Note that the gluon PWs are continuous across $x=\eta$ since ${^g\!\cal P}_j\left(1\right)={^g\!\cal Q}_j\left(1\right)$.

The spin-averaged gluon GPD is captured by the 
symmetric combination  $H_g^{(+)}(x, \eta, t; \mu)$, 
\begin{widetext}
\begin{eqnarray}
\label{ExpansionGPDeigenfunctionsH222GGsymmeticVector}
H_g^{(+)}(x, \eta, t; \mu) = H_g(x, \eta, t; \mu)+H_g(-x, \eta, t; \mu)= \sum_{n=2}^{\infty} (-1)^{n+1} \left( {^g\!p}_n(x, \eta) + {^g\!p}_n(-x, \eta) \right) \mathbb{F}_g^{(+)}(j, \eta, t; \mu),
\end{eqnarray}
where the sum runs over even $n=2,4,\ldots$. Its analytical continuation to complex spin-j, using the Sommerfeld-Watson transform, with support $0\leq x\leq1$, is given by
\begin{eqnarray}
\label{HgFinal}
H_g^{(+)}(x, \eta, t; \mu) = \frac{1}{2i} \int_{\mathbb C} dj\, \frac{(-1)}{\sin(\pi j)} \left( {^g\!p}_j(x, \eta) + {^g\!p}_j(-x, \eta) \right) \mathbb{F}_g^{(+)}(j, \eta, t; \mu),
\end{eqnarray}
\end{widetext}
with the conformal PWs ${^g\!p}_j(x, \eta)$  defined in (\ref{Def-p-allGluonhere}).

\section{Conformal moments of GPDs using the string-based parametrization}
\label{sec_3}
The holographic approach to the space-like form factors in QCD is string based. For a fixed boundary source, the form factor follows from the bulk-to-boundary propagator which naturally 
resums the Reggeized states in bulk. Similarly, the holographic 
description of scattering amplitudes in the Regge limit, is well
described by the resummed spin-j exchanges in bulk, readily capturing the subtle aspects of the multi-parton exchanges in the ERBL regime for  quarks through open strings, and for gluons through closed strings.

In our recent holographic analysis of the GPDs at finite skewness~\cite{Mamo:2022jhp}, we have shown how most of the
GPDs involved in exclusive processes can be obtained holographically in the ERBL regime, where the string based description with its symmetries and dualities, is dominant. 
A string based parametrization for the  conformal moments was derived  in this regime where $|x|<\eta$. The chief idea of this work is to show how this string based parametrization for the conformal spin-j moments can be extended to the DGLAP regime for any skewness, using the analytical continuation of the PWs expansion we detailed above.

\subsection{Quarks conformal moments: singlet-sea}
The holographic parametrization of the conformal moments of the singlet (sea) quark GPDs at $\mu=\mu_0$ is represented by the sum of the sea quark spin-j A-form factor and skewness-dependent spin-j D-form factor as  
\bea\label{seaMomentFinal}
&&\sum_{q=1}^{N_f}\,\mathbb{F}_{q}^{(+)} (j,\eta, t ; \mu_0) =\nonumber\\
&&\sum_{q=1}^{N_f}\,\mathcal{F}_{q}^{(+)}(j,t;\mu_0)+\mathcal{F}_{q\eta}^{(+)}(j,\eta,t;\mu_0)\,,
\eea
for even $j=2,4,\cdots$, where we have defined
\bea\label{singletF1jj}
&&\sum_{q=1}^{N_f}\,\mathcal{F}_{q}^{(+)}(j,t;\mu_0) = \int_{0}^{1}\,dx\,\sum_{q=1}^{N_f}
\frac{q^{(+)}(x;\mu_0)}{x^{1-j+\alpha^{\prime}_{(+)}t}}\,,\nonumber\\ 
\eea
with the input singlet quark PDF 
\bea
&&\sum_{q=1}^{N_f}\,q^{(+)}(x;\mu_0)=\sum_{q=1}^{N_f}\,q(x;\mu_0)+\bar q(x;\mu_0)\nonumber\\
&&=\sum_{q=1}^{N_f}\,H^{(+)}_q (x, \eta=0, t=0 ; \mu_0)
\eea
at $\mu=\mu_0$.
The skewness or
$\eta$-dependent 
spin-j D-form factor is given by
\bea\label{singletF1etaKj22}
&&\sum_{q=1}^{N_f}\,\mathcal{F}_{q\eta}^{(+)}(j,\eta,t;\mu_0)=\left(\hat{d}_{j}(\eta,t)-1\right)\nonumber\\
&&\times \left[\sum_{q=1}^{N_f}\,\mathcal{F}_{q}^{(+)}(j,t;\mu_0)-\mathcal{F}_{q,s}^{(+)}(j,t;\mu_0)\right]\,,\nonumber\\
\eea
where
\bea
\label{singletF1etaSj22}
\sum_{q=1}^{N_f}\,\mathcal{F}_{q,s}^{(+)}(j,t;\mu_0)&\equiv&\sum_{q=1}^{N_f}\,\mathcal{F}_{q}^{(+)}(j,t;\mu_0,\alpha^{\prime}_{(+)}\rightarrow \alpha^{\prime}_{(+),s})\,,\nonumber\\
\eea
and
\bea\label{singletetapoly232}
&&\hat{d}_{j}(\eta,t)-1=\nonumber\\
&&\, _2F_1\left(-\frac{j}{2},-\frac{j-1}{2};\frac{1}{2}-j;\frac{4 m_N^2}{-t}\times\eta ^2\right)-1\,.\nonumber\\
\eea
For even integer $j=2,4,\cdots$, we can also rewrite (\ref{singletetapoly232}) as a finite series (or polynomial) in $\eta$ as
\begin{widetext}
\be\label{singletetapoly2322}
\hat{d}_{j}(\eta,t)-1
=\eta^j\times\left[\frac{(\frac{1-j}{2})_{j/2}}{(\frac{1}{2}-j)_{j/2}}\times\left(\frac{4 m_N^2}{t}\right)^{j/2}\right]\,+\,\sum_{n=1}^{(j-2)/2}\,\eta^{2n}\times\left[\binom{j/2}{n}\frac{(\frac{1-j}{2})_n}{(\frac{1}{2}-j)_n}\times\left(\frac{4 m_N^2}{t}\right)^n\right]\,. \nonumber\\
\ee
\end{widetext}
Here $\binom{m}{n}$ is the Binomial coefficient, and $(q)_n=q(q+1)...(q+n-1)$ for $n>0$ with $(q)_0=1$ is the (rising) Pochhammer symbol. Note that the spin-2 A-form factor of the quark gravitational form factor of the proton is given by  
\be
\sum_{q=1}^{N_f}\,A_q(t;\mu_0)=\sum_{q=1}^{N_f}\,\mathcal{F}_{q}^{(+)}(j=2,t;\mu_0)\nonumber\\
\ee
and the D-form factor (or the D-term) of the quark gravitational form factor of the proton is given by 
\be
\eta^2 \sum_{q=1}^{N_f}\,D_q(t;\mu_0)=\sum_{q=1}^{N_f}\,\mathcal{F}_{q\eta}^{(+)}(j=2,\eta,t;\mu_0)\,.
\ee

\subsection{Quarks conformal moments: non-singlet-valence-isovector}
Using dual gravity, the conformal moments of the non-singlet (valence) quark GPDs,
were explicitly derived by us in~\cite{Mamo:2022jhp}. As we noted earlier, their 
functional dependence in $j,\eta$ holds whatever the regime. More specifically, the non-singlet quark conformal moments at the resolution $\mu=\mu_0$ are represented by the valence quark spin-j A-form factor as~\cite{Mamo:2022jhp} 
\bea\label{valenceMomentFinal}
&&\mathbb{F}_{q}^{(-)} (j,\eta, t ; \mu_0) =\mathcal{F}_{q}^{(-)}(j,t;\mu_0)
\,,
\eea
for odd $j=1,3,\cdots$, which is tied to the valence (non-singlet) quark PDF through
\bea\label{F1jj}
\mathcal{F}_{q}^{(-)}(j,t;\mu_0) = \int_{0}^{1}\,dx\,
\frac{q_v(x;\mu_0)}{x^{1-j+\alpha^{\prime}_{(-)}t}}\,,
\eea
with $q_v(x;\mu_0)=q(x;\mu_0)-\bar q(x;\mu_0)=H^{(-)}_q (x, \eta=0, t=0 ; \mu_0)$ at $\mu=\mu_0$. We can also write
\bea\label{F1jj2}
\mathbb{F}_{u\pm d}^{(-)} (j,\eta, t ; \mu_0) &=&\mathcal{F}_{u\pm d}^{(-)}(j,t;\mu_0)\nonumber\\
&=&\int_{0}^{1}\,dx\,
\frac{u_v(x;\mu_0)\pm d_v(x;\mu_0)}{x^{1-j+\alpha^{\prime}_{u\pm d}t}}\,,\nonumber\\
\eea
assuming $\bar{u}(x,\mu_0)=\bar{d}(x,\mu_0)$. 

In (\ref{valenceMomentFinal}) and (\ref{F1jj2}), we have ignored the valence (non-singlet) quark skewness-dependent spin-j D-form factor which is null (since the pressure distribution inside the nucleon, as determined by the D-form factor, is expected to be flavor or electric charge independent).

We can fix $\alpha_{u\pm d}'$, using the experimental Dirac electromagnetic form factor of the proton ($F_{1p}(t)$) and neutron ($F_{1n}(t)$) from their valence combination
\bea
\mathcal{F}_{u+d}^{(-)}(j=1,t;\mu_0)=3\left(F_{1p}(t)+F_{1n}(t)\right)\,,
\eea
and the isovector combination
\bea
\mathcal{F}_{u-d}^{(-)}(j=1,t;\mu_0)=F_{1p}(t)-F_{1n}(t)\,.
\eea
We specifically use the experimental Dirac electromagnetic form factor of the nucleon from~\cite{Mamo:2021jhj}.

Note that for $j=1$ and low-x, the integrands asymptote 
\bea
\frac{q_v(x;\mu_0)}{x^{\alpha_{(-)}^{\prime} t}}\rightarrow 
\bigg(\frac 1x\bigg)^{\alpha_{(-)}(0)+\alpha_{(-)}^{\prime} t}
\eea
which is the expected open-string Regge limit, with
intercept $\alpha_{(-)}(0)$ and slope $\alpha_{(-)}^{\prime}$.

\subsection{Gluon conformal moments}
The holographic parametrization of the conformal moments of the gluon GPD at finite skewness, and at input scale $\mu=\mu_0$ is represented by the sum of the gluon spin-j A-form factor and skewness-dependent spin-j D-form factor as  
\bea\label{gluonMomnetFinal}
\mathbb{F}_g^{(+)} (j,\eta, t ; \mu_0)=\mathcal{A}_g(j, t;\mu_0)+\mathcal{D}_{g\eta}(j,\eta,t;\mu_0)\,,\nonumber\\
\eea 
for even $j=2,4,...$, where we have defined
\bea\label{Ajj}
&&\mathcal{A}_{g}(j,t;\mu_0) = \int_{0}^{1}\,dx\,
\frac{xg(x;\mu_0)}{x^{2-j+\alpha_T^{\prime}t}}\,.
\eea
Again we note that for $j=2$ and low-x, the integrand asymptotes 
\bea
\frac{xg(x;\mu_0)}{x^{\alpha_T^{\prime} t}}\rightarrow 
\bigg(\frac 1x\bigg)^{\alpha_T(0)+\alpha_T^{\prime} t}
\eea
which is the expected closed-string Regge limit, with
intercept $\alpha_T(0)$ and slope $\alpha^{\prime}_{T}(0)$. 

Also note that (\ref{singletF1jj}), (\ref{F1jj}), and (\ref{Ajj}) can be derived holographically using t-channel string exchange in deformed AdS~\cite{Mamo:2022jhp}. However, the result is dependent on the choice of the particular deformation of the AdS background metric, such as a hard-wall or a soft-wall in  AdS space. Here we will fix them using the empirical quark and gluon PDFs.

The skewness or
$\eta$-dependent terms $\mathcal{D}_{\eta}$ are given by
\bea\label{DKj22}
&&\mathcal{D}_{g\eta}(j,\eta,t;\mu_0)=\nonumber\\
&&\left(\hat{d}_{j}(\eta,t)-1\right)\times \left[\mathcal{A}_{g}(j,t;\mu_0)-\mathcal{A}_{gS}(j,t;\mu_0)\right]\,,\nonumber\\
\eea
where
\bea
\label{ASj22}
\mathcal{A}_{gS}(j,t;\mu_0)&\equiv&\mathcal{A}_{g}(j,t;\mu_0,\alpha^{\prime}_T\rightarrow \alpha^{\prime}_S)\,,
\eea
with the Regge slope parameter $\alpha^{\prime}_S$ to be fixed by the gluon gravitational form factor of the proton, and $\hat d_j(\eta, t)-$ given by (\ref{singletetapoly232}) or its finite series expansion 
(\ref{singletetapoly2322}), for  even integer $j=2,4,\cdots$.
Note that the spin-2 A-form factor of the gluonic gravitational form factor of the proton is given by~\cite{Mamo:2022jhp} 
\be
A_g(t;\mu_0)=\mathcal{A}_{g}(j=2,t;\mu_0)
\ee
and the D-form factor (or the D-term) of the gluonic gravitational form factor of the proton is given by 
\be
\eta^2 D_g(t;\mu_0)=\mathcal{D}_{g\eta}(j=2,\eta,t;\mu_0)\,.\nonumber\\
\ee 

We would like to highlight recent findings from the $J/\Psi-007$ collaboration at JLab~\cite{Duran:2022xag}. This collaboration employed the holographic scattering amplitude for near-threshold $J/\Psi$ production by us~\cite{Mamo:2019mka,Mamo:2022eui}, which correlates with the spin-2 moment $\mathbb{F}_g^{(+)} (2,\eta, t ; \mu_0)$ as defined in (\ref{gluonMomnetFinal}). Through their experiments, they successfully extracted the gluonic spin-2 moment, which show strong agreement with results from lattice QCD~\cite{Pefkou:2021fni,Hackett:2023rif}. This supports our holographic ansatz for the conformal moments as outlined in (\ref{gluonMomnetFinal}), establishing a robust experimental and theoretical foundation.

\begin{figure*}
\centering
\subfloat[\label{FUminusDn1}]{%
\includegraphics[height=6cm,width=.48\linewidth]{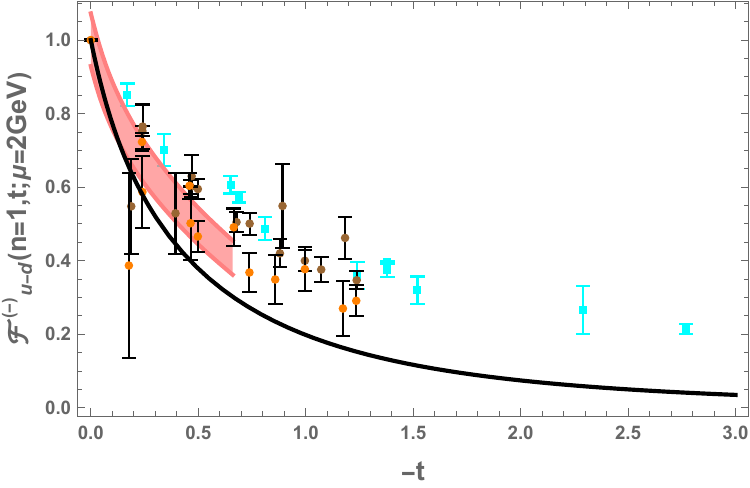}%
}\hfill
\subfloat[\label{FUminusDn2}]{%
\includegraphics[height=6cm,width=.48\linewidth]{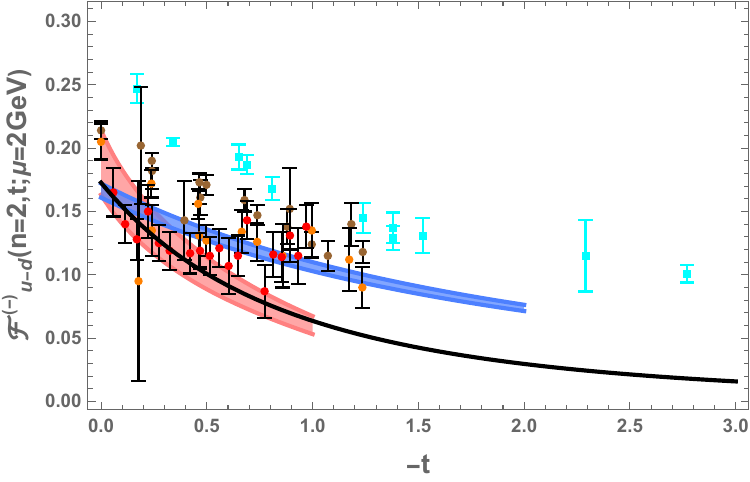}%
}\hfill
\subfloat[\label{FUminusDn3}]{%
\includegraphics[height=6cm,width=.48\linewidth]{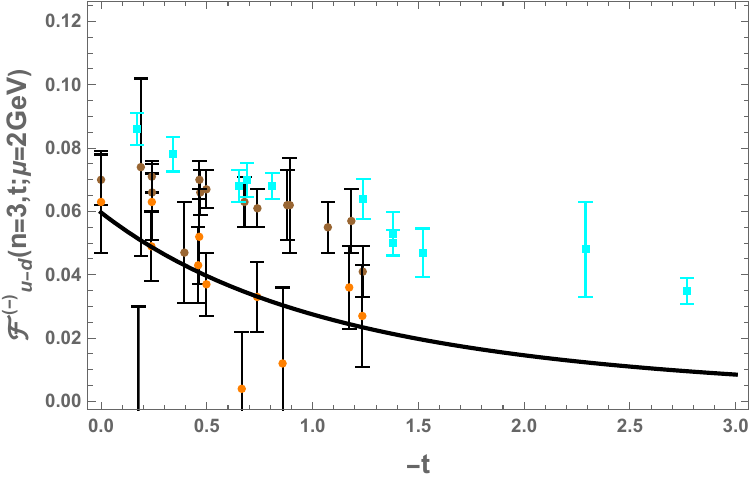}%
}\hfill
\subfloat[\label{FUminusDn4}]{%
\includegraphics[height=6cm,width=.48\linewidth]{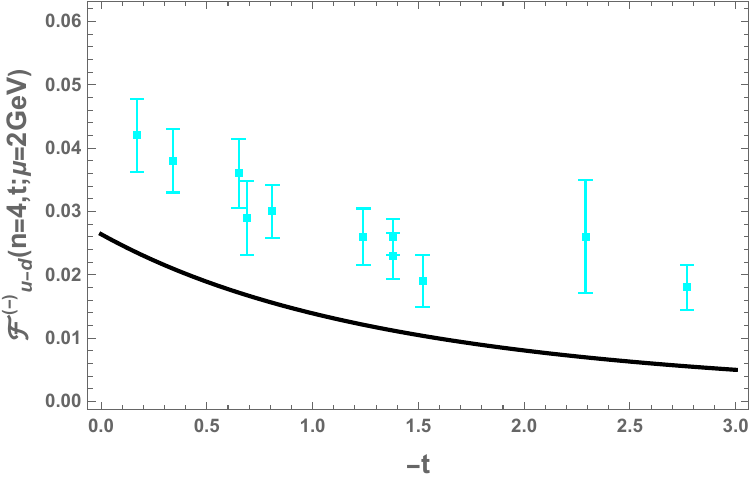}%
}
\hfill
\subfloat[\label{FUminusDn5}]{%
\includegraphics[height=6cm,width=.48\linewidth]{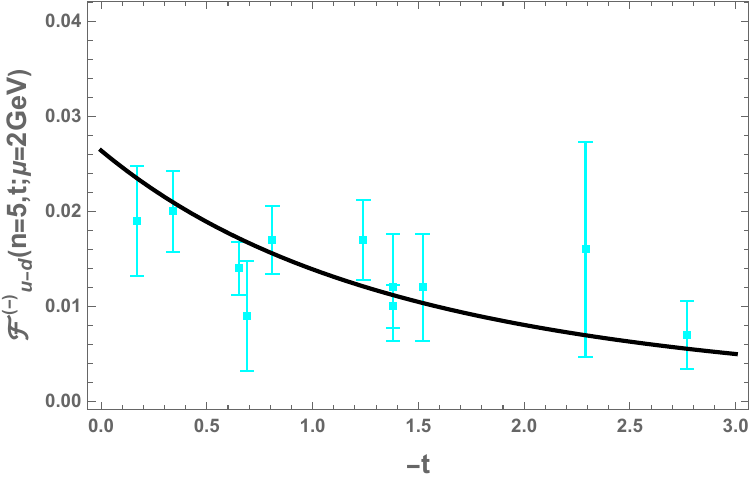}%
}
\caption{Our evolved moments of $u-d$ quark GPD $H^{u-d}(x,\eta,t;\mu)$ at $\mu=2~\rm GeV$ (black-solid line). The other colored curves and data points
are lattice resukts from~\cite{Bhattacharya:2023ays} (Cyan data points), \cite{LHPC:2007blg} (Orange and Brown data points), \cite{Lin:2020rxa} (Pink curve), \cite{Alexandrou:2019ali} (Red data points), and \cite{Hackett:2023rif} (Blue curve) are shown for comparison.}
\label{HUMINUSDmoments}
\end{figure*}

\begin{figure*}
\centering
\subfloat[\label{FUplusDn1}]{%
\includegraphics[height=6cm,width=.48\linewidth]{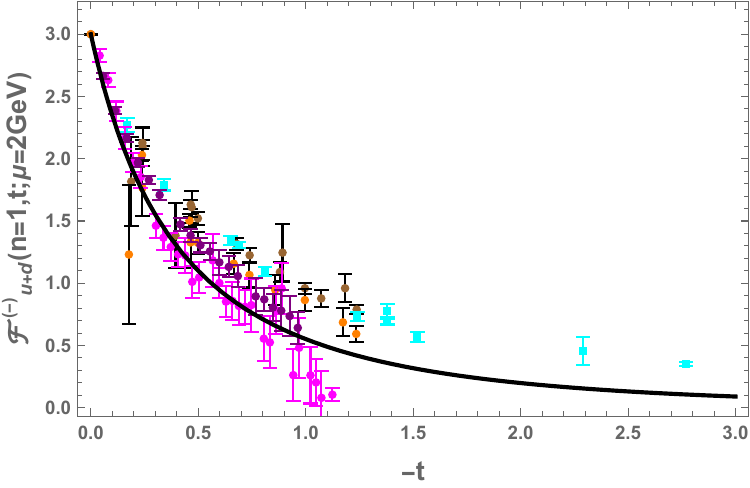}%
}\hfill
\subfloat[\label{FUplusDn2}]{%
\includegraphics[height=6cm,width=.48\linewidth]{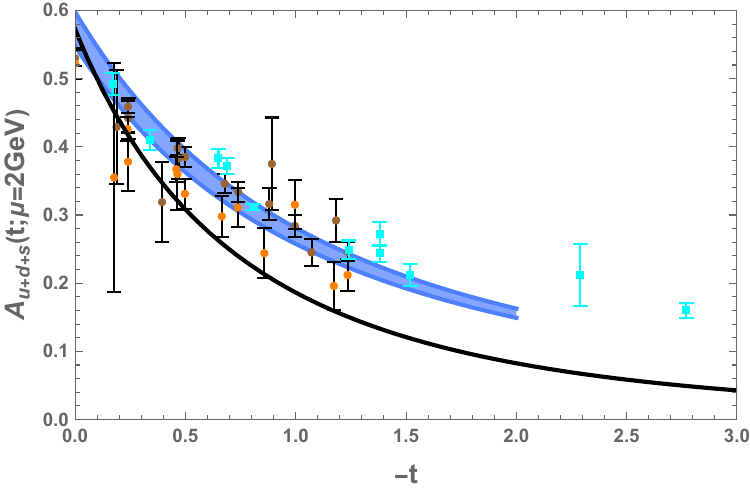}%
}\hfill
\subfloat[\label{DUplusDplusSn2}]{%
\includegraphics[height=6cm,width=.48\linewidth]{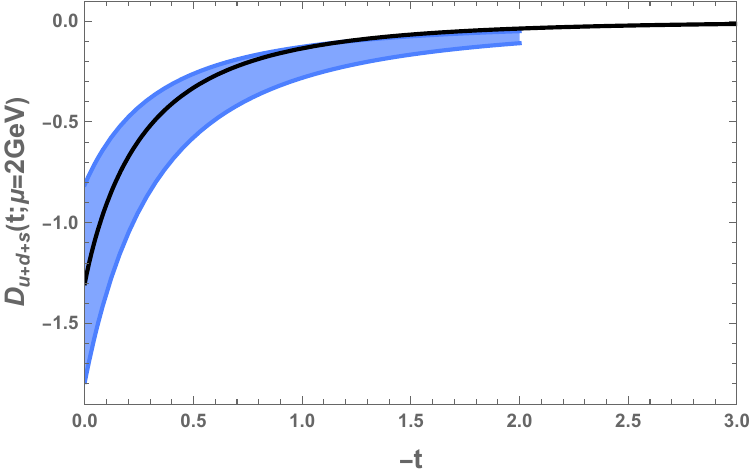}%
}\hfill
\subfloat[\label{FUplusDn3}]{%
\includegraphics[height=6cm,width=.48\linewidth]{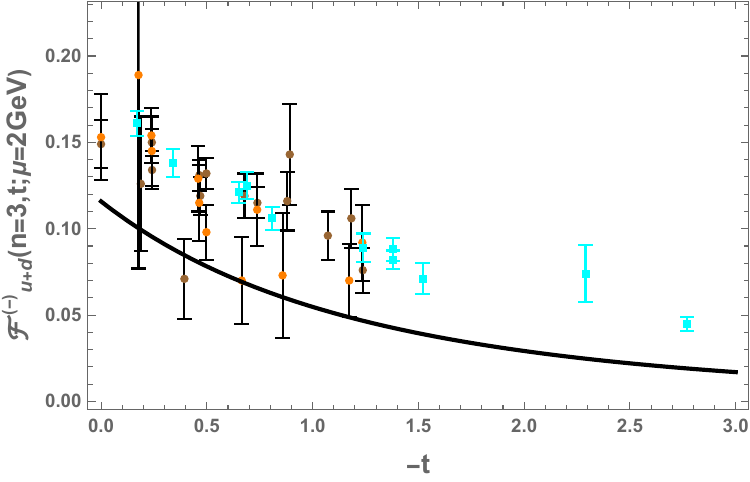}%
}
\hfill
\subfloat[\label{FUplusDn4}]{%
\includegraphics[height=6cm,width=.48\linewidth]{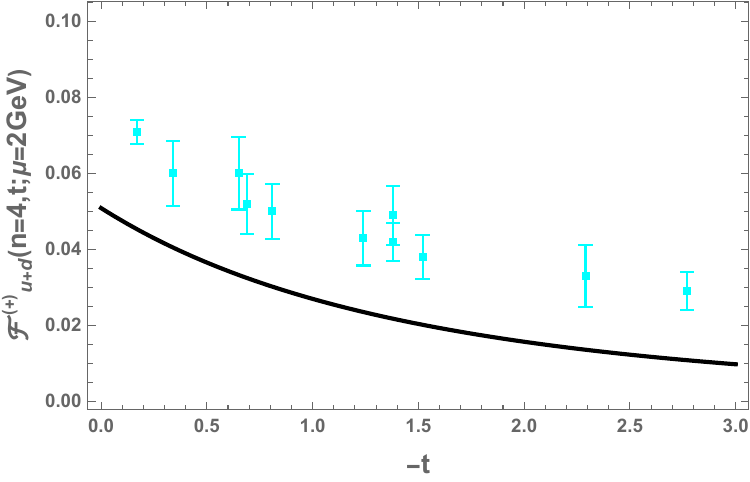}%
}\hfill
\subfloat[\label{FUplusDn5}]{%
\includegraphics[height=6cm,width=.48\linewidth]{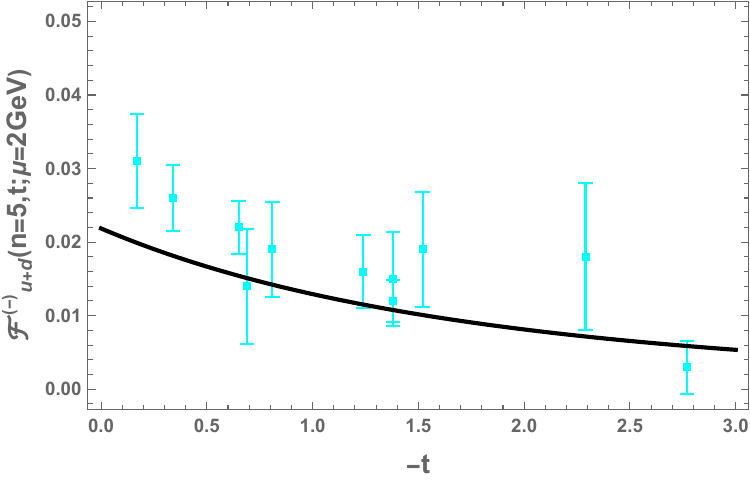}%
}
\caption{Our evolved moments of $u+d$ quark GPD $H^{u+d}(x,\eta,t;\mu)$ at $\mu=2~\rm GeV$ (black-solid line). The other colored Conformal Regge theory are lattice results from~\cite{Bhattacharya:2023ays} (Cyan data points), \cite{LHPC:2007blg} (Orange and brown data points), \cite{Alexandrou:2018sjm} (Purple data points), \cite{Djukanovic:2023beb} (Magenta data points), and \cite{Hackett:2023rif} (Blue curve) are shown for comparison.}
\label{HUPLUSDDmoments}
\end{figure*}

\begin{figure*}
\centering
\subfloat[\label{AgluonN2}]{%
\includegraphics[height=6cm,width=.48\linewidth]{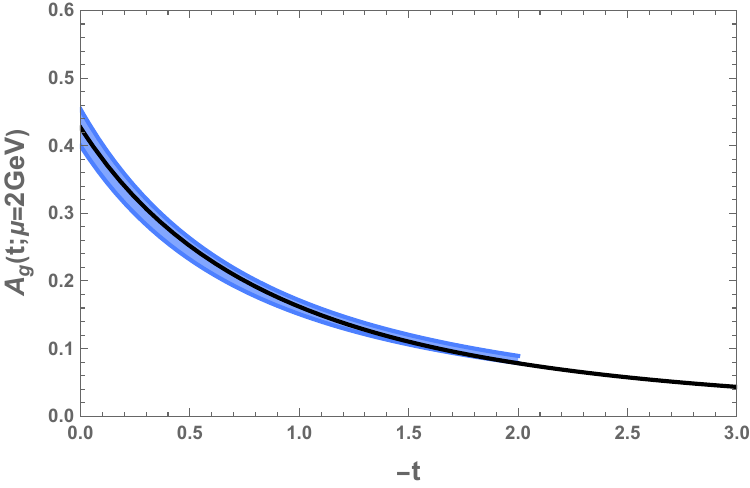}%
}\hfill
\subfloat[\label{DgluonN2}]{%
\includegraphics[height=6cm,width=.48\linewidth]{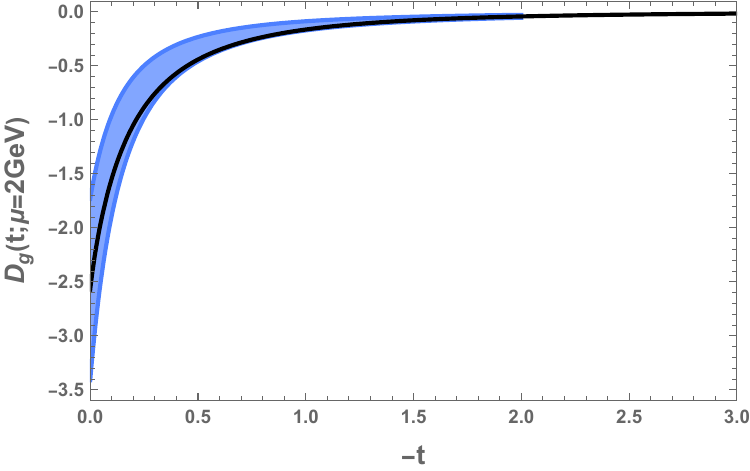}%
}
\caption{Our evolved moments of the symmetric gluon GPD $H^{(+)}_{g}(x,\eta,t;\mu)$ at $\mu=2~\rm GeV$, are represented by black line. The blue curve corresponding to lattice data from \cite{Hackett:2023rif} is shown for comparison.}
\label{HGLUONmoments}
\end{figure*}

\subsection{Evolution of conformal moments}
\label{EVO}
The leading-order evolution of the conformal moments for valence (non-singlet) quark GPDs is given as follows~\cite{Mamo:2022jhp}
\begin{equation}\label{NSevolution}
\mathbb{F}_q^{(-)} (j,\eta, t ; \mu) = \mathbb{F}_q^{(-)} (j,\eta, t ; \mu_0) \left( \frac{\alpha_s (\mu_0^2)}{\alpha_s (\mu^2)} \right)^{\frac{\gamma_{j-1}^{qq;\text{NS}}}{\beta_0}}\,.
\end{equation}
Here, $\gamma_{j}^{qq;\text{NS}}\equiv\gamma_{(0),j}^{qq; \text{V};\text{NS}}$ denotes the leading-order (LO) vector non-singlet anomalous dimensions:
\bea
&&\gamma_{j}^{qq;\text{NS}} =\nonumber\\
&&- C_F \left( - 4 \psi(j + 2) + 4 \psi(1) + \frac{2}{(j + 1)(j + 2)} + 3 \right),\nonumber\\
\eea
with $\psi(z)$ representing the digamma function and $\gamma_{\scriptscriptstyle\rm E} = - \psi (1) \approx 0.577216$.

Note that the non-singlet isovector conformal moments are evolved as
\begin{equation}\label{NSevolution}
\mathbb{F}_{u-d}^{(-)} (j,\eta, t ; \mu) = \mathbb{F}_{u-d}^{(-)} (j,\eta, t ; \mu_0) \left( \frac{\alpha_s (\mu_0^2)}{\alpha_s (\mu^2)} \right)^{\frac{\gamma_{j-1}^{qq;\text{NS}}}{\beta_0}}\,.
\end{equation}

The evolution equations for the singlet quark and symmetric gluon GPD conformal moments are provided by~\cite{Mamo:2022jhp}
\begin{align}\label{Sevolution}
&\sum_{q=1}^{N_f} \mathbb{F}_q^{(+)}(j,\eta, t ; \mu) =\left(\frac{\gamma_{j-1}^{qq}}{\gamma_{j-1}^{+} - \gamma_{j-1}^{-}}\right) \nonumber\\
&\times\Bigg[\Gamma^{-}_{j-1}\,\mathbb{F}^+_j(j,\eta, t;\mu_0)\left( \frac{\alpha_s (\mu^2_0)}{\alpha_s (\mu^2)} \right)^{\frac{\gamma_{j-1}^{+}}{\beta_0}}\nonumber\\
& -\Gamma^{+}_{j-1}\mathbb{F}^-_j(j,\eta, t;\mu_0)\left( \frac{\alpha_s (\mu^2_0)}{\alpha_s (\mu^2)} \right)^{\frac{\gamma_{j-1}^{-}}{\beta_0}}\Bigg],
\end{align}
\begin{align}
&\mathbb{F}_g^{(+)}(j,\eta, t ; \mu) = \left(\frac{\frac{6}{j}\gamma_{j-1}^{gq}}{\gamma_{j-1}^{+} - \gamma_{j-1}^{-}}\right)\nonumber\\
&\times\Bigg[\mathbb{F}^+_j(j,\eta, t;\mu_0)\left( \frac{\alpha_s (\mu^2_0)}{\alpha_s (\mu^2)} \right)^{\frac{\gamma_{j-1}^{+}}{\beta_0}}\nonumber\\
& -\mathbb{F}^-_j(j,\eta, t;\mu_0)\left( \frac{\alpha_s (\mu^2_0)}{\alpha_s (\mu^2)} \right)^{\frac{\gamma_{j-1}^{-}}{\beta_0}}\Bigg],
\end{align}
where $\Gamma^{\pm}_j\equiv 1-\frac{\gamma_{j}^{\pm}}{\gamma_{j}^{qq}}$,
\begin{align}
&\mathbb{F}^{\pm}_j(j,\eta, t ; \mu_0) = \sum_{q=1}^{N_f} \mathbb{F}_q^{(+)}(j,\eta, t ; \mu_0)\nonumber\\
&+ \left(\frac{\frac{j}{6}\gamma_{j-1}^{qg}}{\gamma_{j-1}^{qq} - \gamma_{j-1}^{\mp}}\right) \mathbb{F}_g^{(+)}(j,\eta, t ; \mu_0),
\end{align}
and the eigenvalues $\gamma^{\pm}_j$ are defined as:
\begin{equation}
\gamma^{\pm}_j = \frac{1}{2}\left( \gamma_{j}^{qq} + \gamma_{j}^{gg} \pm \sqrt{(\gamma_{j}^{qq} - \gamma_{j}^{gg})^2 + 4\gamma_{j}^{gq}\gamma_{j}^{qg}} \right).
\end{equation}
The leading-order (LO) vector singlet anomalous dimensions $\gamma^{ab}_{j}\equiv\gamma^{ab;V}_{(0)j}$ (with $a=q,g$ and $b=q,g$) are given in~\cite{Mamo:2022jhp} (see  Eqs.K.31-38), for both the axial-vector (A) and vector (V) ones. Here we need the vector (V) ones.

\section{Results}
\label{sec_4}
To proceed, we need to fix the initial quark and gluon conformal moments in
(\ref{F1jj},\ref{singletF1jj},\ref{Ajj}). Since the holographic predictions for these moments 
rely on the choice of the bulk metric, we suggest
to use the empirical quark and gluon  PDFs at the
lowest resolution point $\mu=\mu_0$, for a model independent initial assessment. In addition, the quark and gluon Regge slopes $\alpha^{\prime}_{u\pm d,T}$ will
be treated as free parameters, to be fixed empirically. 

More specifically, we will use the leading-order Martin-Stirling-Thorne-Watt (MSTW) 2008 PDF sets~\cite{Martin:2009iq} at $\mu_0=1~\text{GeV}$,
as quoted in Appendix~\ref{app_1} for completeness. Their pertinent  evolutions to higher resolution say $\mu=2\, {\rm GeV}$, 
are summarized in subsection~\ref{EVO}. Although
the Regge slopes are free parameters  fixed empirically, the canonical string values are $\alpha^{\prime}_{q}=\left(\alpha^{\prime}_{u+d}+\alpha^{\prime}_{u-d}\right)/2\approx 2\alpha^{\prime}_T=2\alpha^{\prime}_g\approx 1\,{\rm GeV}^{-2}$ for comparison.  

Our study leverages the holographic parametrization, relying on only the quark and gluon empirical Regge slope parameters, listed in Table~\ref{table1}, $\alpha^{\prime}_{u\pm d}$ (including $\alpha^{\prime}_{(+),s}$) and $\alpha^{\prime}_{T,S}$ are fixed by the electromagnetic and gravitational form factors of the proton.  We note that 
the empirical quark and gluon Regge intercepts fixed
by the global fits, are consistent with the canonical string values. 
 
We now proceed to compare
in details our results for the holographically parametrized GPDs, with the recently reported lattice nucleon quark GPDs, both for the lowest moments and  also the full quark GPDs at finite skewness.


\subsection{Moments versus lattice}

In Fig.~\ref{HUMINUSDmoments}, we compare our evolved lowest moments $n=1,2,..,5$ for the nucleon valence $u-d$ (isovectror) quark GPD  $H^{u-d}(x,\eta,t;\mu)\equiv H_{u-d}^{(-)}(x,\eta,t;\mu)$ at $\mu=2~\rm GeV$, to the recently reported lattice data. Our results are
represented by the black-solid lines. The lattice results from different collaborations are shown as
data points or colored spreads. The lattice results are  from~\cite{Bhattacharya:2023ays} (Cyan data points), \cite{LHPC:2007blg} (Orange and Brown data points), \cite{Lin:2020rxa} (Pink curve), \cite{Alexandrou:2019ali} (Red data points), and \cite{Hackett:2023rif} (Blue curve).
Overall, our results are in relative good agreement with the reported lattice results.

In Fig.~\ref{HUPLUSDDmoments} we compare our  evolved lowest moments for the nucleon sea $u+d$ quark GPD $H^{u+d}(x,\eta,t;\mu)$ at $\mu=2~\rm GeV$, to the lattice results. Again,  our results are
represented by the black-solid lines. The lattice results from different collaborations are shown as
data points or colored spreads. They are 
from~\cite{Bhattacharya:2023ays} (Cyan data points), \cite{LHPC:2007blg} (Orange and Brown data points), \cite{Alexandrou:2018sjm} (Purple data points), \cite{Djukanovic:2023beb} (Magenta data points), and \cite{Hackett:2023rif} (Blue curve) are shown for comparison. Again, the agreement with the reported lattice results, is relatively good.

In Fig.~\ref{HGLUONmoments} we compare our evolved and symmetric gluon  GPD $H^{(+)}_{g}(x,\eta,t;\mu)$ at $\mu=2~\rm GeV$, to the lattice results reported in~\cite{Hackett:2023rif}. Our results are 
represented by the black-solid line, and the lattice results by the blue-spread. The agreement 
of the holographic results with the lattice results is good.

\begin{figure*}
\centering
\subfloat[\label{HUminusDoursAll}]{%
\includegraphics[height=6cm,width=.48\linewidth]{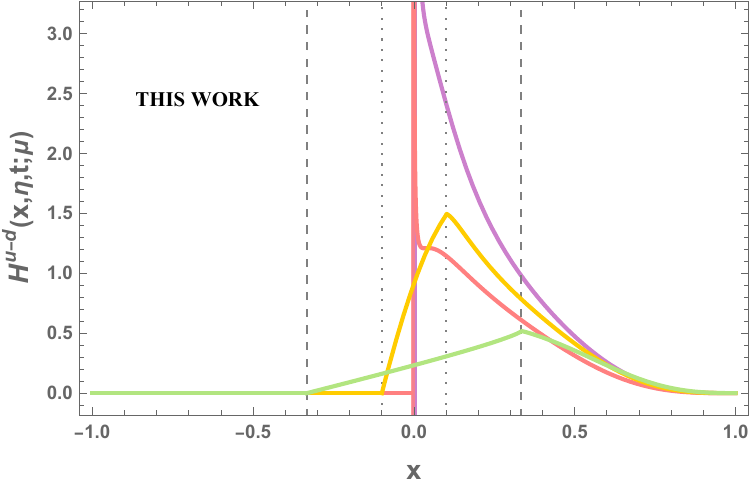}%
}\hfill
\subfloat[\label{HUminusDLatticeMarthaLinAll}]{%
\includegraphics[height=6cm,width=.48\linewidth]{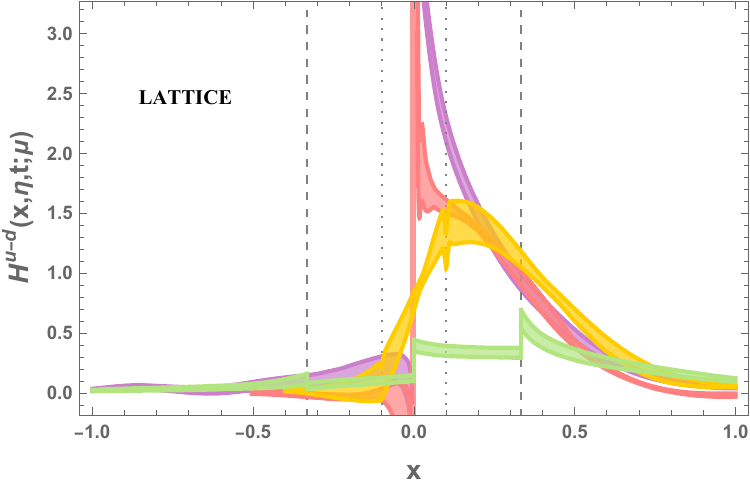}%
}
\caption{(a) The colored solid curves represent our numerical results for the $u-d$ (isovector) quark GPD $H^{u-d}(x,\eta,t;\mu)$ of (\ref{umdExpansionGPDValenceVector}) using the conformal moments (\ref{F1jj2}), interpolated from our numerical data. The curves correspond to various kinematic configurations: purple for $\eta=0$, $-t=0$, and $\mu=2~\rm GeV$, green for $\eta=1/3$, $-t=0.69~\rm GeV^2$, and $\mu=2~\rm GeV$, yellow for $\eta=0.1$, $-t=0.23~\rm GeV^2$, and $\mu=2~\rm GeV$, and pink for $\eta=0$, $-t=0.39~\rm GeV^2$, and $\mu=3~\rm GeV$. (b) Corresponding lattice results from~\cite{Alexandrou:2020zbe} (purple, and green curves), \cite{Holligan:2023jqh} (yellow curve), and~\cite{Lin:2020rxa} (pink curve) are shown with matching color coding for direct comparison.}
\label{HUMINUSDGPDs2}
\end{figure*}

\begin{figure*}
\centering
\subfloat[\label{HUminusDeta0MinusT0}]{%
\includegraphics[height=6cm,width=.48\linewidth]{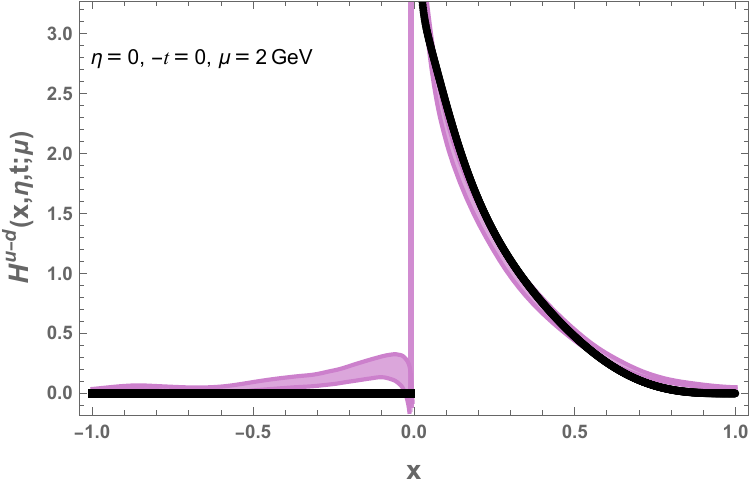}%
}\hfill
\subfloat[\label{HUminusDeta0MinusT0pt69}]{%
\includegraphics[height=6cm,width=.48\linewidth]{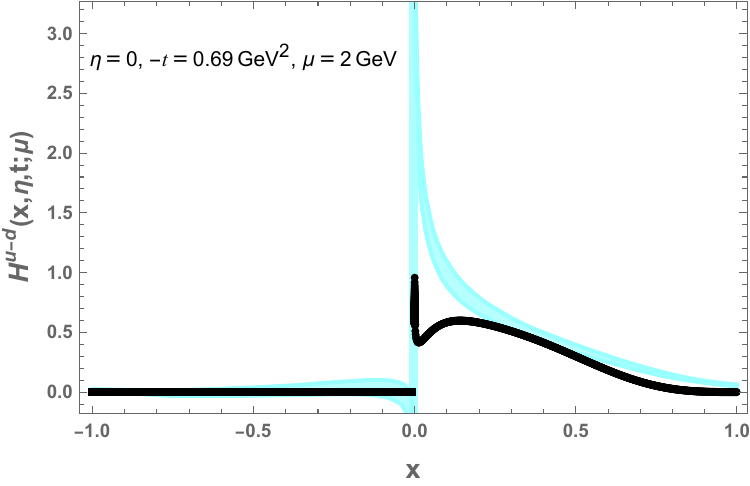}%
}\hfill
\subfloat[\label{HUminusDeta0MinusT0pt39Mu3}]{%
\includegraphics[height=6cm,width=.48\linewidth]{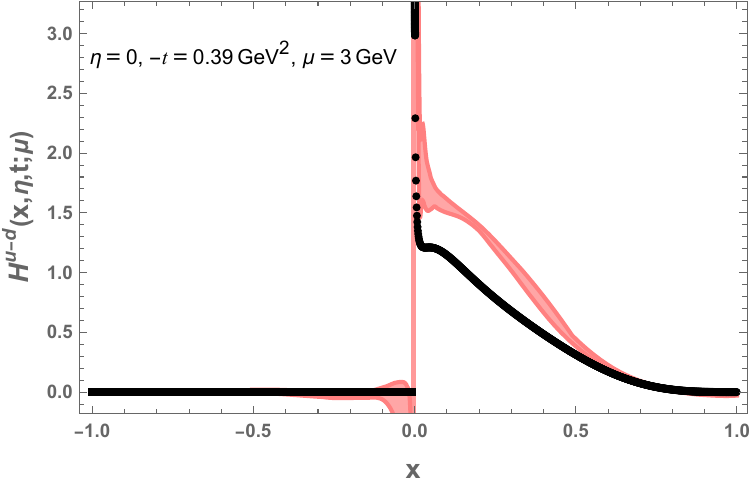}%
}\hfill
\subfloat[\label{HUminusDeta1over3MinusT0pt69}]{%
\includegraphics[height=6cm,width=.48\linewidth]{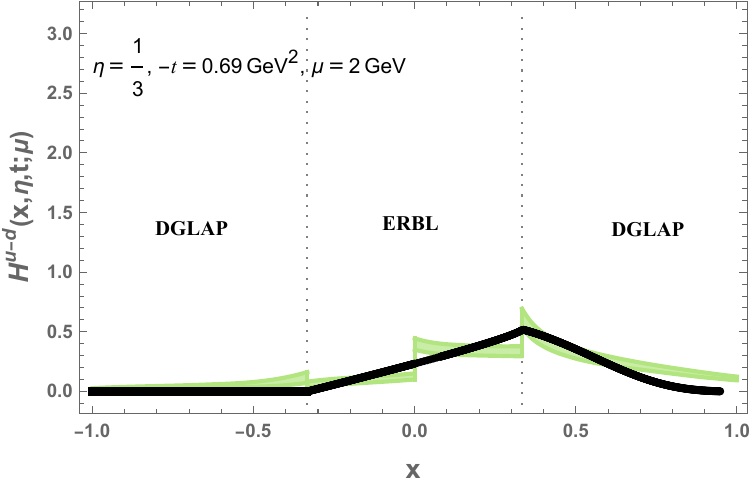}%
}
\hfill
\subfloat[\label{HUminusDeta0pt1MinusT0pt23}]{%
\includegraphics[height=6cm,width=.48\linewidth]{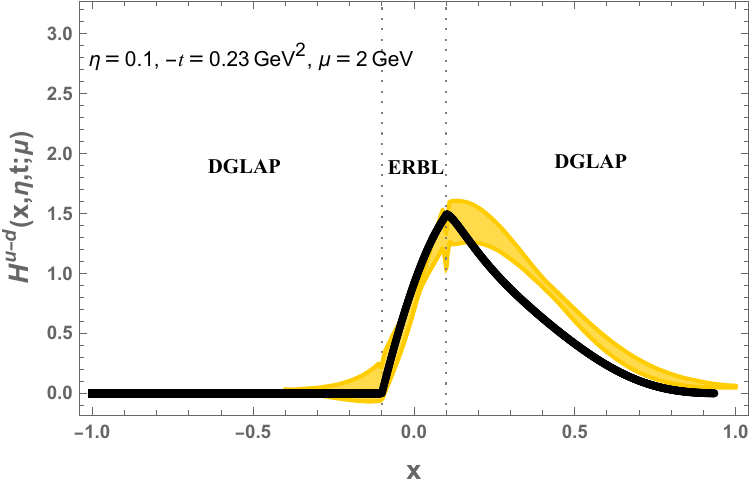}%
}
\caption{Our numerical results, for the $u-d$ (isovector) quark GPD $H^{u-d}(x,\eta,t;\mu)$, are represented by black data points. The other colored curves, corresponding to lattice data, from~\cite{Alexandrou:2020zbe} (Purple, Cyan, and Green curves), \cite{Holligan:2023jqh} (Blue curve), and~\cite{Lin:2020rxa} (Pink curve) are shown for comparison. 
The figures illustrate various kinematic scenarios: (\textbf{a}) $\eta=0$, $-t=0$, and $\mu=2~\rm GeV$ (\textbf{b}) $\eta=0$, $-t=0.69~\rm GeV^2$, and $\mu=2~\rm GeV$ (\textbf{c}) $\eta=0$, $-t=0.39~\rm GeV^2$, and $\mu=3~\rm GeV$, (\textbf{d}) $\eta=1/3$, $-t=0.69~\rm GeV^2$, and $\mu=2~\rm GeV$ (\textbf{e}) $\eta=0.1$, $-t=0.23~\rm GeV^2$, and $\mu=2~\rm GeV$.}
\label{HUMINUSDGPDs}
\end{figure*}

\begin{figure*}
\centering
\subfloat[\label{HuAll}]{%
\includegraphics[height=6cm,width=.48\linewidth]{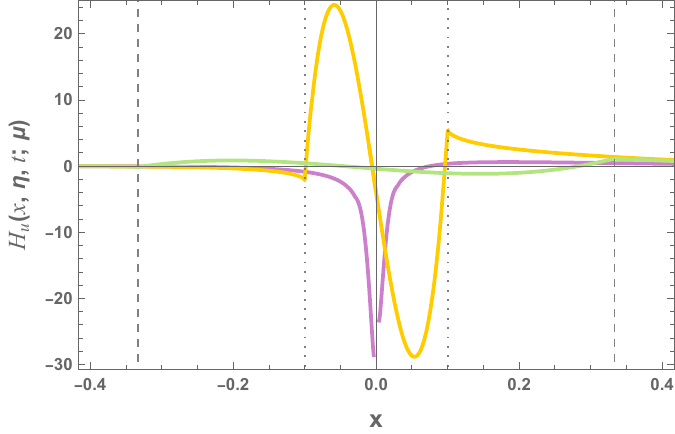}%
}\hfill
\subfloat[\label{HdAll}]{%
\includegraphics[height=6cm,width=.48\linewidth]{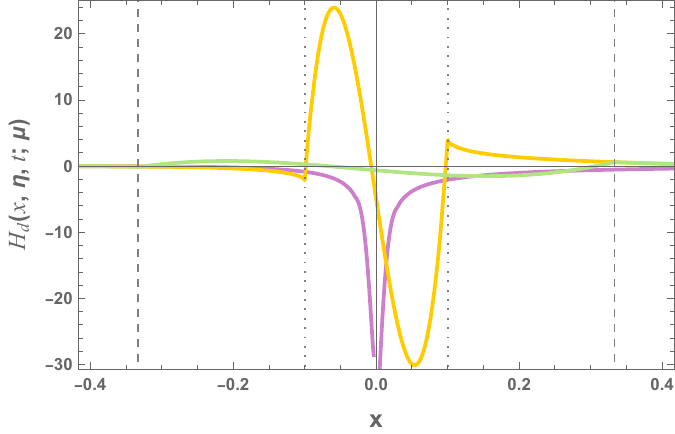}%
}
\caption{The colored solid curves represent our numerical results for the flavor separated $u$ and $d$ quark GPDs $H^{u,d}(x,\eta,t;\mu)$, interpolated from our numerical data. The curves refer to various kinematics: Purple for $\eta=0$, $-t=0$, and $\mu=2~\rm GeV$, Green for $\eta=1/3$, $-t=0.69~\rm GeV^2$, and $\mu=2~\rm GeV$, and Yellow for $\eta=0.1$, $-t=0.23~\rm GeV^2$, and $\mu=2~\rm GeV$.}
\label{HUD}
\end{figure*}

\begin{figure*}
\centering
\subfloat[\label{gluonHallData}]{%
\includegraphics[height=6cm,width=.48\linewidth]{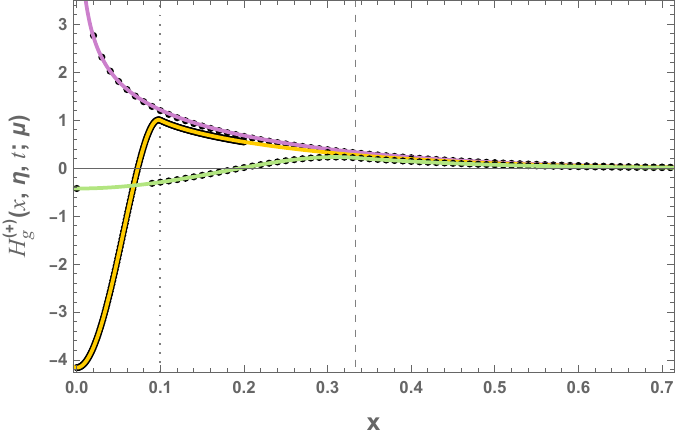}%
}\hfill
\subfloat[\label{seaHallData}]{%
\includegraphics[height=6cm,width=.48\linewidth]{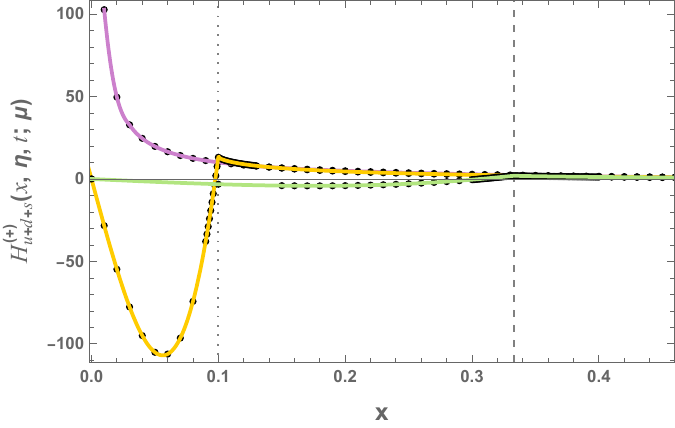}%
}\hfill
\subfloat[\label{uValenceHallData}]{%
\includegraphics[height=6cm,width=.48\linewidth]{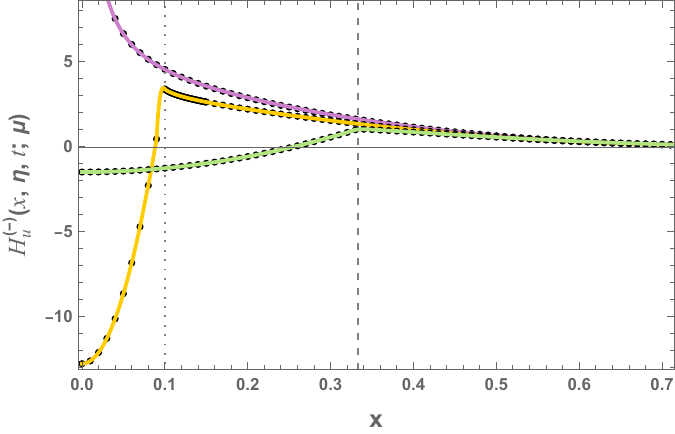}%
}\hfill
\subfloat[\label{dValenceHallData}]{%
\includegraphics[height=6cm,width=.48\linewidth]{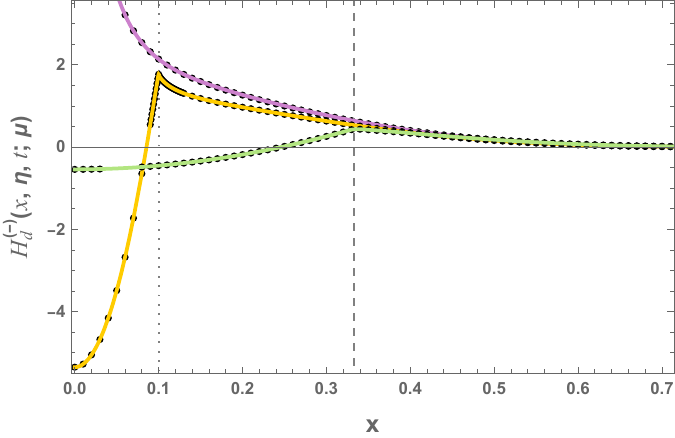}%
}
\caption{Our numerical results for the gluon GPD (a), sea quark GPD (b), valence up quark GPD (c), and valence down quark GPD (d) are represented by black data points. The colored curves are our interpolation of the numerical data. The figures illustrate various kinematic scenarios: (Purple curves) $\eta=0$, $-t=0$, and $\mu=2~\rm GeV$, (Yellow curves) $\eta=0.1$, $-t=0.23~\rm GeV^2$, and $\mu=2~\rm GeV$, (Green curves) $\eta=1/3$, $-t=0.69~\rm GeV^2$. All curves are evolved to the resolution  $\mu=2~\rm GeV$.}
\label{Hgluonseavalencedata}
\end{figure*}

\subsection{GPDs versus lattice}

In Fig.~\ref{HUMINUSDGPDs2}a we show our full GPDs for the nucleon valence $u-d$ (isovector) quark GPD $H^{u-d}(x,\eta,t;\mu)$ of (\ref{umdExpansionGPDValenceVector}) using the conformal moments (\ref{F1jj2}), for different values of the skewness $\eta$ and
Mandelstam $t$.  The colored curves refer to different kinematics with Purple for $\eta=0$, $-t=0$, and $\mu=2~\rm GeV$, Green for $\eta=1/3$, $-t=0.69~\rm GeV^2$, and $\mu=2~\rm GeV$, Yellow for $\eta=0.1$, $-t=0.23~\rm GeV^2$, and $\mu=2~\rm GeV$, and Pink for $\eta=0$, $-t=0.39~\rm GeV^2$, and $\mu=3~\rm GeV$. 
In Fig.~\ref{HUMINUSDGPDs2}b the corresponding lattice results with colored curves are
shown for comparison. The lattice results are from~\cite{Alexandrou:2020zbe} (Purple, and Green curves), \cite{Holligan:2023jqh} (Yellow curve), and~\cite{Lin:2020rxa} (Pink curve) are shown with matching color coding for direct comparison.

In Fig.~\ref{HUMINUSDGPDs}, we show a one-to-one comparison of our results in black-solid lines, for the nucleon valence $u-d$ quark GPD $H^{u-d}(x,\eta,t;\mu)$, for different kinematics. The
colored curves are the lattice data from~\cite{Alexandrou:2020zbe} (Purple, Cyan, and Green curves), from~\cite{Holligan:2023jqh} (Blue curve), and from~\cite{Lin:2020rxa} (Pink curve) 
with
\\
\indent {\bf a:}\,\,\,  $\eta=0$, $-t=0$, and $\mu=2~\rm GeV$; \\
\indent {\bf b:}\,\,\, $\eta=0$, $-t=0.69~\rm GeV^2$, and $\mu=2~\rm GeV$;   \\
\indent {\bf c:}\,\,\, $\eta=0$, $-t=0.39~\rm GeV^2$, and $\mu=3~\rm GeV$;\\
\indent {\bf d:}\,\,\, $\eta=1/3$, $-t=0.69~\rm GeV^2$, and $\mu=2~\rm GeV$;\\
\indent {\bf e:}\,\,\, $\eta=0.1$, $-t=0.23~\rm GeV^2$, and $\mu=2~\rm GeV$.
\\
\noindent The overall agrrement with the reported lattice results at different kinematical regimes, is again good.

\subsection{Gluon, valence and sea quark GPDs}

In Fig.~\ref{gluonHallData} we show our prediction for the gluon GPD $H_g^{(+)}(x,\eta,t;\mu)$, as defined in (\ref{HgFinal}). The prediction follows from the gluon conformal moments from (\ref{gluonMomnetFinal}), based on the MSTW2008lo gluon PDF from \cite{Martin:2009iq}. Our explicit results are shown in open circles with numerical interpolation for different kinematics $\eta=0, -t=0$
(purple curve), $\eta=0.1, -t=0.23\, \rm GeV^2$ (yellow curve) and 
$\eta=1/3, -t=0.69\,\rm GeV^2$ at the evolved resolution $\mu=2\, \rm GeV$.

In Fig.~\ref{seaHallData}, we show our predictions for the singlet (sea) quark GPD $H_{u+d+s}^{(+)}(x,\eta,t;\mu)$, following from (\ref{HseaFinal}). The results follow from the singlet quark conformal moments in (\ref{seaMomentFinal}) and the MSTW2008lo singlet (sea) quark PDF in~\cite{Martin:2009iq}. In addition, we use the Regge slope parameters $\alpha^{\prime}_{(+)}=\alpha^{\prime}_{u+d}$, and $\alpha^{\prime}_{(+),s}$ listed in Table~\ref{table1}. The curve color
display and kinematics is the same as that in Fig.~\ref{gluonHallData}.

For the non-singlet (valence) quark GPDs, Figs.~\ref{uValenceHallData}-\ref{dValenceHallData} show our predictions for $H_{u,d}^{(-)}(x,\eta,t;\mu)$, 
as in (\ref{HvalenceFinal}). The results follow from the non-singlet quark conformal moments given in (\ref{valenceMomentFinal}), using the MSTW2008lo non-singlet (valence) quark PDF \cite{Martin:2009iq}. The curve color
display and kinematics is the same as that in Fig.~\ref{gluonHallData}.

\subsection{Flavor separated quark GPDs}
In Fig.~\ref{HUD} we show our predictions for the flavor separated nucleon $u,d$ quark GPDs $H^{u,d}(x,\eta,t;\mu)$. For the flavor separation, we have ignored the strange quark contribution and assumed $H_{u+d+s}^{(+)}\approx H_{u+d}^{(+)}$. The flavor separated GPDs reduce to the expected $u,d$ PDFs at $\eta=t=0$ with the expected differences (Purple curves). The differences are less pronounced for the $u,d$ GPDs at $\eta=1,3, -t=0.69\,\rm GeV^2$
(Green curves) and $\eta=0.1, -t=0.23\,\rm GeV^2$ (Yellow curves).
For the flavor (up (u) quark and down (d) quark) separation, as well as for separation into quark and anti-quark (in the positive (+x), and negative (-x) regions of quark GPDs, respectively), we made use of the following relations
\begin{widetext}
\bea
&&H^{u+d}(+x,\eta,t;\mu)=\frac{1}{2}\left(H_{u+d}^{(-)}(x,\eta,t;\mu)+H_{u+d}^{(+)}(x,\eta,t;\mu)\right)\,,\\
&&H^{u+d}(-x,\eta,t;\mu)=\frac{1}{2}\left(H_{u+d}^{(-)}(x,\eta,t;\mu)-H_{u+d}^{(+)}(x,\eta,t;\mu)\right)\,,\\
&&H^{u}(+x,\eta,t;\mu)=\frac{1}{2}\left(H^{u+d}(+x,\eta,t;\mu)+H^{u-d}(+x,\eta,t;\mu)\right)\,,\\
&&H^{u}(-x,\eta,t;\mu)=\frac{1}{2}\left(H^{u+d}(-x,\eta,t;\mu)+H^{u-d}(-x,\eta,t;\mu)\right)\,,\\
&&H^{d}(+x,\eta,t;\mu)=\frac{1}{2}\left(H^{u+d}(+x,\eta,t;\mu)-H^{u-d}(+x,\eta,t;\mu)\right)\,,\\
&&H^{d}(-x,\eta,t;\mu)=\frac{1}{2}\left(H^{u+d}(-x,\eta,t;\mu)-H^{u-d}(-x,\eta,t;\mu)\right)\,.
\eea
\end{widetext}

\begin{table}
\caption{Regge slope parameters for the holographic parametrization of conformal moments.}
\label{table1}
\centering
\begin{tabular}{cc}
\hline \hline
Parameter & Value ($\rm GeV^{-2}$) \\
\hline
$\alpha^{\prime}_T$ & 0.627 \\
$\alpha^{\prime}_S$ & 4.277 \\
$\alpha^{\prime}_{(+)}=\alpha^{\prime}_{(-)}=\alpha^{\prime}_{u+d}$ & 0.891 \\
$\alpha^{\prime}_{u-d}$ & 1.069 \\
$\alpha^{\prime}_{(+),s}$ & 1.828 \\
\hline \hline
\end{tabular}
\end{table}

\section{Conclusions}
\label{sec_5}
The string-based parametrization  we have presented, shows how to extend our recent proposal for the nucleon GPDs~\cite{Mamo:2022jhp} from the ERBL regime at low-x, to the DGLAP regime at large-x. More specifically, the GPDs are expanded 
in terms of  integer valued conformal moments, using an ortho-complete  basis set of conformal PWs. Using the Sommerfeld-Watson transform, the series is then analytically continued in the complex j-plane, wth a pertinent analytical continuation of the PWs whatever the skewness.  This proposal is general, and holds for any
dynamical perturbative or non-perturbative analysis of the 
GPDs. 

The spin-j conformal moments in the integral transform, encode the essential hadronic content of the GPDs. In the gravity dual approach at large skewness and small parton-x, they 
are described by Reggeized exchanges of bulk spin-j mesons (open string Regge) or spin-j glueballs (closed string Regge). These exchanges are captured by
generalized string based  form factors through pertinent confluent hypergeometric functions~\cite{Mamo:2022jhp}, with only the 
Regge intercepts and slopes as parameters. 
Analyticity as enforced by the integral transforms, allows for the string based result
to be used as parametrizations of the spin-j
conformal moments for any skewness.

At zero skewness, the spin-j conformal moments are moments of the pertinent quark and gluon  parton distribution functions. At low resolution, we have elected to fix these spin-j conformal moments using the empirical MSTW PDFs, since the gravity dual description of  the PDFs is mostly limited to the Regge regime. Their evolution to
higher resolution follows from general QCD evolution, as they are form invariant. Overall, our results for the  string based parametrization of the GPDs for the moments of the nucleon singlet and non-singlet quark GPDs are in fair agreement  with those reported by various lattice collaborations. We expect more precise results with physical quark masses from lattice will be in better agreement with our moments. Also our results for the nucleon isovector quark GPDs at different kinematics with zero and finite skewness, are in
good agreement with most of the lattice results reported recently. More importantly, we have used the string based parametrization to predict the nucleon gluon,
valence and sea quark GPDs, which are yet to be accessed by current lattice simulations. The extension of our analysis to the polarized, and helicity flip 
nucleon GPDs, as well as other hadrons will be addressed in upcoming publications.

In summary, the string based parametrization of the quark and gluon  GPDs we have presented, obey all the constraints required by symmetry and analyticity. Their explicit form
should prove useful for  the program of global GPDs data analyses, bypassing the de-convolution problem, to be carried out by the quark-gluon-tomography (QGT) collaboration. 
Most importantly, they should also be useful for the analysis of a number of exclusive processes currently carried out at JLab, and planned for the
upcoming EIC.

\vskip 1cm
\centerline{\bf Acknowledgments}
\vskip 0.5cm
K.M. is supported by U.S. DOE Grant No. DE-FG02-04ER41302, and thanks Konstantinos Orginos and Christian Weiss for discussions and hospitality at JLab where part of this work was carried out. I.Z. is supported by the  U.S. DOE Grant  No. DE-FG-88ER40388.
This research is also supported in part within the framework of the Quark-Gluon Tomography (QGT) Topical Collaboration, under contract no. DE-SC0023646.

\appendix

\section{MSTW PDFs}
\label{app_1}
To fix the input conformal moments of the quark and gluon GPDs at low resolution $\mu=\mu_0$, we 
will use the empirical nucleon quark and gluon PDFs, instead of the holgraphic PDFs which depend on the choice of the bulk metric, and hence model dependent. More specifically, we will use 
the leading-order Martin-Stirling-Thorne-Watt (MSTW) 2008 PDF sets~\cite{Martin:2009iq} at $\mu_0=1~\text{GeV}$
\begin{widetext}
\bea
u_v(x;\mu_0)\equiv u(x;\mu_0)-\bar{u}(x;\mu_0)&=&1.4335\, x^{-0.54768}(1-x)^{3.0409} \left(8.9924 x-2.3737 \sqrt{x}+1\right)\,,\nonumber\\
d_v(x;\mu_0)\equiv d(x;\mu_0)-\bar{d}(x;\mu_0)&=&5.0903\, x^{-0.28022} (1-x)^{5.1244} \left(7.473 x-4.3654 \sqrt{x}+1\right)\,,\nonumber\\
s_v(x;\mu_0)\equiv s(x;\mu_0)-\bar{s}(x;\mu_0)&=&0.011523\, x^{-1.299} (1-x)^{10.285} \left(\frac{x}{0.017414}-1\right)\,,\nonumber\\
%
%
\Delta(x;\mu_0)\equiv\bar{d}(x;\mu_0)-\bar{u}(x;\mu_0)&=&8.9413\,x^{0.876} (1-x)^{10.8801} \left(-36.507 x^2+8.4703 x+1\right) \,,\nonumber\\
s^{(+)}(x;\mu_0)\equiv s(x;\mu_0)+\bar{s}(x;\mu_0)&=&0.10302\,x^{-1.16276} (1-x)^{13.242} \left(16.865 x-2.9012 \sqrt{x}+1\right)\,,\nonumber\\
xg(x;\mu_0)&=&0.0012216\,x^{-0.83657} (1-x)^{2.3882} \left(1445.5 x-38.997 \sqrt{x}+1\right)\,
\eea
and
\bea
S(x;\mu_0)&\equiv& 2\left(\bar{u}(x;\mu_0)+\bar{d}(x;\mu_0)\right)+s(x;\mu_0)+\bar{s}(x;\mu_0)\nonumber\\
&=&0.59964\,x^{-1.16276} (1-x)^{8.8801} \left(16.865 x-2.9012 \sqrt{x}+1\right)\,.
\eea
We note that in our numerical evaluation of the sea and $u-d$ (isovector) quark GPDs, we used
\bea
\sum_{q=1}^{N_f=3}\,H^{(+)}_q (x, \eta=0, t=0 ; \mu_0)=\sum_{q=1}^{N_f=3}\,q(x;\mu_0)+\bar q(x;\mu_0)=u_v(x;\mu_0)+d_v(x;\mu_0)+S(x;\mu_0)\,,
\eea
and
\bea
H^{u-d}(x,\eta,t;\mu_0)\approx u_v(x;\mu_0)-d_v(x;\mu_0)\,,
\eea
respectively, with $\Delta(x;\mu_0)=\bar{d}(x;\mu_0)-\bar{u}(x;\mu_0)\approx 0$.

\end{widetext}

\bibliography{Parametrized_GPDs_vs_lattice_GPDs}

\end{document}